\documentstyle[11pt]{article}
\begin{document}
\begin{large}
\vskip 5mm
\noindent

\title{Probing flavor changing interactions in hadron collisions
\footnote{This work was supported by National Natural Science Foundation of
          China.}}

\author{{
  Chang Chao-Hsi$^{a,b}$, Han Liang$^{c}$, Jiang Yi$^{c}$, Ma Wen-Gan$^{a,b,c}$,
  Zhou Hong$^{c}$, Zhou Mian-Lai$^{c}$ }\\
{\small $^{a}$CCAST (World Laboratory), P.O.Box 8730, Beijing 100080,
China.}\\
{\small $^{b}$Institute of Theoretical Physics, Academia Sinica,}\\
{\small P.O.Box 2735, Beijing 100080, China.} \\
{\small $^{c}$Department of Modern Physics, University of Science
and Technology}\\
{\small of China (USTC), Hefei, Anhui 230027, China. } }
\date{}
\maketitle
\begin{center}\begin{minipage}{100mm}

\vskip 5mm
\begin{center} {\bf Abstract}\end{center}
\baselineskip 0.3in
{The subprocess $gg \rightarrow t\bar{c}+\bar{t}c$ in the two-Higgs-doublet
model with flavor-changing scalar couplings is examined at the one loop
level. With perturbative QCD factorization theorem, the corresponding
cross sections for hadron-hadron collisions are computed numerically. The
results are applicable to the whole mass range of the weakly coupled Higgs
bosons. In case we could efficiently exclude the severe backgrounds of the
$t\bar{c}(\bar{t}c)$ production signal, probing the flavor-changing
top-charm-scalar vertex at hadron colliders would be very promising and
accessible experimentally.
} \\

\vskip 5mm
{PACS number(s):13.85.Ni, ~11.30.Hv, ~12.60.Fr, ~14.65.Ha}
\end{minipage}
\end{center}

\newpage
\noindent
{\Large{\bf I. Introduction}}
\par
There are stringent experimental constraints against the existence
of tree level neutral flavor changing interactions, especially for light
quarks\cite{str}. This leads to the naturalness to suppress the flavor
changing neutral current interactions (FCNCs) for all kinds of model
building studies, that is realized in terms of the Glashow-Iliopoulos-Maiani
(GIM) mechanism in the standard model(SM).
\par
The two-Higgs-doublet models(THDMs) are extensions of the SM in which one
more scalar doublet is added. In order to forbid possible tree level FCNCs
to appear in the models, Glashow and Weinberg\cite{NFC} proposed a
neutral flavor conservation (NFC) condition by imposing discrete symmetries
on the models. The common THDMs with the NFC condition can be divided into
two categories, i.e., Model I and Model II. In Model I, both the up- and
down-type quarks couple to the same one of the Higgs doublets, but in the
Model II the up- and down-type quarks couple to the two Higgs doublets
respectively. In fact, the suppression facts of the  FCNC processes in the
cases involving the light down-type quark sector have been observed
experimentally\cite{exp}, whereas those relevant to the up-type quark sector
have not been established so well. Therefore the so-called Model-III of the
THDM is proposed and it allows the FCNC Yukawa couplings which has the
character that the couplings are related to the masses of the coupled flavors
even at tree level\cite{Luke}.
\par
As pointed out by Cheng, Sher and other authors\cite{fs}\cite{bfs}
\cite{chengs}\cite{M3}, the Yukawa couplings are typically proportional to
the masses of the coupled fermions at the vertices, it is rather natural to
imagine having such Yukawa couplings for the FCNC interactions instead of
placing the constraints on the theory. In this case, low energy limits on
the FCNCs may be evaded, because the flavor changing couplings to the light
quarks are small, and the suppression of the FCNC processes involving light
quarks can be automatically satisfied. Then the imposition of discrete
symmetries to obtain the NFC condition, therefore, is unnecessary, which
is normally invoked in common THDMs to prevent the FCNC interactions at
tree level.  In the literature such the THDM without the NFC condition is
called as the `third' THDM (i.e., THDM III) \cite{M3}, where the up-type and
down-type quarks are allowed simultaneously to couple to more than one scalar
doublet. In the framework of the model the effects of the FCNC interactions
involving the heavy quark will be enhanced.
\par
Now the top quark has been observed and a unexpected very large mass
$m_{t} = 173.8 \pm 5.2~GeV$ (world average value) has been
recognized\cite{CDF}. This `extraordinary' mass scale of the top
quark may have many important implications pertaining to many outstanding
issues. One of them is to test the FCNC processes requested in the Model-III.
Namely, it is a `good place' to observe the facts of the Model-III for the
FCNC interactions. The measurement of FCNC processes involving
top quark would provide an important test for the discrimination of
various models. In the THDM III one would expect that large effects
of the FCNC interactions could manifest themselves in the cases involving the
massive top quark. Therefore testing the existence of the flavor changing
scalar interactions involving top quarks are clearly important.
D. Atwood et al. \cite{atwood} presented results of a calculation for the
process $e^{+}e^{-}\rightarrow t \bar{c}$(or $\bar{t}c$)in the THDM III,
and they obtained $R^{tc}/\lambda^{4}$ to be in the order of $10^{-5}$ with
proper parameters. Recently, Jiang et al.\cite{jiang}studied the production
rates of $e^{+}e^{-} \rightarrow \gamma \gamma \rightarrow t\bar{c}+\bar{t}c$
for the NLC and found that this process is more promising than the straight
production via $e^{+}e^{-}$ collisions for probing the FCNC interactions.
Abraham et al.\cite{abraham} also investigated the anomalous $\bar{t}q\gamma$
couplings via single top quark production processes by considering effective
Lagrangian of the lowest dimension with $\gamma\gamma$ collisions,
and found the processes would be observable with suitable strength of
anomalous coupling as long as b-tagging and suitable kinematic cuts
are taken properly.
\par
After the termination of the running of the LEP2, the hadron colliders
Tevatron and LHC will be the only machines in searching for the FCNC
processes. It is believed that more experimental events involving top quark
will be collected in these hadronic machines. It will give a good chance
to study the physics relevant to the FCNC processes of the Model-III.
In this paper we are to study the problem and present complete one-loop
calculation for the subprocess $gg \rightarrow t\bar{c}$(or $\bar{t}c$) to
the order $O(m_{t}m_{c}/m_{W}^{2})$ for the THDM Model-III. In fact, the
obtained results in the paper are applicable to the whole mass range for
weakly coupled Higgs bosons. The production cross sections of $pp(\bar p)
\rightarrow gg \rightarrow t\bar{c}+\bar{t}c+X$ are also given for the Tevatron
and LHC. The paper is organized as follows: The details of the calculation
are given in Sec. II. In Sec. III numerical results, discussions
and a short summary are presented. Finally, the explicit expressions
used in the paper are collected in appendix.

\vskip 5mm
{\Large{\bf II. Calculation}}
\vskip 5mm
\par
In the third type of the two-Higgs-doublet model, the up-type and down-type
quarks are allowed simultaneously to couple to more than one scalar doublet.
Since there is no global symmetry that distinguishes the two doublets in
the model, we will assume that only one of them $(\phi_{1})$ develops a
vacuum expectation value and the second one $(\phi_{2})$ remains unbroken, i.e.
$$
<\phi_{1}>=
\left(
\begin{array}{l}
0 \\
v/\sqrt{2}
\end{array}
\right) ,
~~<\phi_{2}>=0
\eqno{(1)}
$$
where $v \simeq 246~GeV$. The physical spectrum of Higgs bosons consists of two
scalar neutral bosons $h^{0}$ and $H^{0}$, one pseudoscalar neutral boson
$A^{0}$ and two charged Higgs $H^{\pm}$,
$$
\begin{array}{l}
H^{0}=\sqrt{2}[(Re\phi_{1}^{0}-v)\cos{\alpha}+Re\phi_{2}^{0}\sin\alpha],\\
h^{0}=\sqrt{2}[-(Re\phi_{1}^{0}-v)\sin{\alpha}+Re\phi_{2}^{0}\cos\alpha],\\
A^{0}=\sqrt{2}(-Im\phi_{2}^{0}).
\end{array}
\eqno{(2)}
$$
The masses of the five neutral and charged Higgs bosons and the mixing angle
$\alpha$ are free parameters of the model. The Yukawa couplings to quarks
are\cite{Luke},
$$
{\cal L}^{Q}_{Y}=\lambda^{U}_{ij}\bar{Q_{i}}\tilde{\phi_{1}}U_{j}+
                \lambda^{D}_{ij}\bar{Q_{i}}\phi_{1}D_{j}+
                \xi^{U}_{ij}\bar{Q_{i}}\tilde{\phi_{2}}U_{j}+
                \xi^{D}_{ij}\bar{Q_{i}}\phi_{2}D_{j},
\eqno{(3)}
$$
where the first two terms give masses of the quark mass
eigenstates, and $\xi^{U}_{ij}$ and $\xi^{D}_{ij}$ are the $3\times 3$
matrices which give the strength of the flavor changing neutral scalar
vertices. The $\xi$s are all free parameters and can be constrained by the
experimental data. If we neglect CP violation, the $\xi$s are all
real. We will use the Cheng-Sher Ansatz(CSA)\cite{chengs} and let
$$
\xi_{ij}\sim \frac{\sqrt{m_{i}m_{j}}}{v}.
$$
And we can parametrize the Yukawa couplings as
$$
\xi_{ij}=g \frac{\sqrt{m_{i}m_{j}}}{m_{W}} \lambda.
\eqno{(4)}
$$
Comparing it with the usual gauge couplings of $SU(2)\times U(1)$,
one has $\lambda=\frac{1}{\sqrt{2}}$. In our calculation we use
$\lambda=\frac{1}{\sqrt{2}}$ and note that there is no theoretical
bound on the coupling factor $\lambda$.
\par
The subprocess producing $t\bar{c}(\bar{t}c)$ via gluon-gluon collisions,
$$
gg \rightarrow t\bar{c}(\bar{t}c)
$$
can be induced through one-loop diagrams at the lowest order, and the Feynman
diagrams are drawn in figure 1(a) and figure 1(b), where the contributions
from the one-loop diagrams involving neutral and charged Higgs bosons are
presented, respectively. The contributions from the figures of Fig.1(b) with
the charged Higgs boson in loops being replaced by W-boson, is much smaller
due to the GIM suppression and Yukawa coupling. We can omitted this part
in our calculation. The diagrams exchanging the two external gluon-gluon
lines are not shown, but are numbered in Fig.1(a) and Fig.1(b).
Fig.1(a)(1 $\sim$ 12) and Fig. 1(b)(1 $\sim$ 6) are the self-energy diagrams,
Fig. 1(a)(13 $\sim$ 20) and Fig. 1(b)(7 $\sim$ 10) are the vertex correction
diagrams, Fig. 1(a)(25 $\sim$ 34) and Fig. 1(b) (13 $\sim$ 15) are the
s-channel diagrams, Fig. 1(a)(21 $\sim$ 24) and Fig. 1(b)(11 $\sim$ 12)
are the box diagrams. Note that in the present case at one-loop level the
ultraviolet divergence would be canceled automatically, if all the one-loop
diagrams at the $O(m_{t} m_{c}/m_{W}^{2})$ order in the THDM III are included.
In this work, we perform the calculation in the t'Hooft-Feynman gauge.
\par
To simplify the calculation we set $\alpha=0$ and let all scalar bosons be
degenerate, i.e., $m_{h^{0}}=m_{A^{0}}=m_{H^{\pm}}=M_{s}$ where $M_{s}$ is
the common scalar mass. The contribution from the coupling involving $H^{0}$
is suppressed due to $\alpha=0$.
\par
In the calculation for the s-channel diagrams(Fig.1.(a)(25 $\sim$ 28)), we
take into account the width effects of the $h^{0}$ and $A^{0}$ propagators.
The decays of $h^{0}$ to WW and ZZ are suppressed, because
of the factor $\sin\alpha$ in the $h^{0}WW$ and $h^{0}ZZ$ couplings,
and $h^{0}$ decay to $A^{0}A^{0}$ is also forbidden due to the case of
the degenerate masses of $h^{0}$ and $A^{0}$. Note that the pseudoscalar
$A^{0}$ does not couple with gauge boson pair. Therefore only the
decays of $h^{0}$ and $A^{0}$ to final states $q_{i}\bar{q_{j}}$ need
to be considered, where $q_{i}$ and $q_{j}$ represent quarks of
flavor i and j respectively. The decay width for the scalar $h^{0}$ can
be written as\cite{width}
$$
\Gamma (h^{0}\rightarrow q\bar{q})=\frac{3 g^{2} m_{h^{0}}}{32\pi M_{W}^{2}}
m_{q}\left(1-4m_{q}^{2}/m_{h^{0}}^{2}\right)^{3/2}
$$
and
$$
\Gamma (h^{0}\rightarrow t\bar{c}+\bar{t}c)=\frac{3g^{2}m_{h^{0}}}
{32\pi M_{W}^{2}}\cdot 2m_{t}m_{c}\left[1-(m_{t}+m_{c})^{2}/m_{h^{0}}^{2}
\right]^{3/2}\times \left[1-(m_{t}-m_{c})^{2}/m_{h^{0}}^{2}\right]^{1/2}.
\eqno{(5)}
$$
The decay width for the pseudoscalar $A^{0}$ boson can be represented
by exchanging exponents $3/2 \leftrightarrow 1/2$
and $m_{h^{0}} \leftrightarrow m_{A^{0}}$ in Eq.(5). When
$m_{t} + m_{c} < M_{s} < 2 m_{t}$, the dominant decay modes of $h^{0}$ and
$A^{0}$ are $h^{0}, A^{0} \rightarrow c\bar{c}, b\bar{b},t\bar{c}+\bar{t}c$,
whereas when $M_{s} > 2m_{t}$, the final state $t\bar{t}$ decay channel
is open, and their decay widths are rather large due to the large
masses of $M_{s}$ and $m_{t}$.
\par
We denote $\theta$ as the scattering angle between one of the gluons and
the final top quark. Then we express all the four-momenta of the initial
and final particles in the center-of-mass(CMS) by means of the total
energy $\sqrt{\hat{s}}$ and the scattering angle $\theta$. The four-momenta
of top quark and charm quark are $p_{1}$ and $p_{2}$ respectively and are
read
$$\begin{array}{l}
p_{1}=\left(E_{t}, \sqrt{E_{t}^{2}-m_{t}^{2}}sin\theta, 0,
\sqrt{E_{t}^{2}-m_{t}^{2}}cos\theta\right),\\
p_{2}=\left(E_{c}, -\sqrt{E_{c}^{2}-m_{c}^{2}}sin\theta, 0,
-\sqrt{E_{c}^{2}-m_{c}^{2}}cos\theta\right),
\end{array}
\eqno{(6)}
$$
where
$$
E_{t}=\frac{1}{2}\left(\sqrt{\hat{s}}+(m_{t}^{2}-m_{c}^{2})/\sqrt{\hat{s}}
\right),~~~~~
E_{c}=\frac{1}{2}\left(\sqrt{\hat{s}}-(m_{t}^{2}-m_{c}^{2})/\sqrt{\hat{s}}
\right).
\eqno{(7)}
$$
$p_{3}$ and $p_{4}$ are the four-momenta of the initial gluons and are
expressed as
$$
p_{3}=\left(\frac{1}{2}\sqrt{\hat{s}}, 0, 0, \frac{1}{2}\sqrt{\hat{s}}\right),
~~~~
p_{4}=\left(\frac{1}{2}\sqrt{\hat{s}}, 0, 0, -\frac{1}{2}\sqrt{\hat{s}}\right).
\eqno{(8)}
$$
\par
The corresponding matrix element for all the diagrams in figure 1(a) and
figure 1(b) is written as
$$
M = Tr(T^{a}T^{b})\delta_{jl}M^{\hat{s}_{1}}+(f_{abc}T^{c}_{jl})M^{\hat{s}_{2}}+
    (T^{a}_{jm}T^{b}_{ml}) M^{\hat{t}}+(T^{b}_{jm}T^{a}_{ml})M^{\hat{u}}
\eqno{(9)}
$$
\par
The upper indexes  $\hat{s}_{1}$, $\hat{s}_{2}$, $\hat{t}$ and $\hat{u}$
represent the amplitudes corresponding to the s-channel diagrams Fig. 1(a)
(25$\sim$ 28), s-channel diagrams Fig. 1(a) (29 $\sim$ 34) and
Fig. 1(b) (13 $\sim$ 15), t-channel and u-channel diagrams in figure 1(a)
and figure 1(b) respectively. The $T^{a} (a=1-8)$ are the $SU(3)_{c}$
generators introduced by Gell-Mann and $f_{abc}$ are the antisymmetric
$SU(3)_{c}$ structure constants. The subscripts $j,l (j, l=1,2,3)$ of $T^{a}$
represent the color of final state top quark and charm quark respectively.
The variables $\hat{s}$, $\hat{t}$
and $\hat{u}$ are usual Mandelstam variables in the center of mass system of
gluon-gluon. Their definitions are:
$$
\begin{array}{l}
\hat{s}=(p_{1}+p_{2})^{2}=(p_{3}+p_{4})^{2},~~~~
\hat{t}=(p_{1}-p_{3})^{2}=(p_{2}-p_{4})^{2},\\
\hat{u}=(p_{1}-p_{4})^{2}=(p_{2}-p_{3})^{2}.
\end{array}
\eqno{(10)}
$$
\par
We collect all the explicit expressions of the amplitudes appearing in
equation (9) in the appendix. The total cross section for
$g g \rightarrow t\bar{c}+\bar{t}c$ can be written in the form
$$
\hat{\sigma}(\hat{s})=\frac{2}{16\pi \hat{s}^2}
\int_{\hat{t^{-}}}^{\hat{t^{+}}} d\hat{t} \vert \bar{M}\vert^{2}
\eqno{(11)}
$$
where $\vert \bar{M}\vert^{2}$ is the initial spin-averaged matrix element
squared  and
$\hat{t^{\pm}}=1/2(m_{t}^{2}+m_{c}^{2}-\hat{s})\pm \sqrt{E_{t}^{2}-m_{t}^{2}}
\sqrt{\hat{s}}$.
The cross section for $ p p \rightarrow g g \rightarrow
t\bar{c}+\bar{t}c +X$ is conveniently written in terms of the rapidities
$y_{1}$ and $y_{2}$ of the two jets (finial states) and their common
transverse momentum $p_{T}$. Here we neglect the intrinsic transverse
momentum carried by partons. It is
$$
  \frac{d\sigma}{dy_{1}dy_{2}dp_{T}}=\frac{\pi \tau p_{T}}{\hat{s}}
  f_{g}(x_1,Q^{2}) f_{g}(x_2,Q^{2}) \hat{\sigma}  (gg \rightarrow
t\bar{c}+\bar{t}c~ {\rm at} ~ \hat{s}=\tau s), \eqno(12)
$$
where $\sqrt{s}$ and $\sqrt{\hat{s}}$ denote the proton-proton
and gluon-gluon c.m. energies respectively and $\hat{s}=s\tau$.
$f_{g}(x_i,Q^{2})$ is the distribution function of gluon in proton.
\par
Defining
$$
y^{*}=\frac{1}{2}(y_{1}-y_{2}) \eqno(13)
$$
and
$$
y_{boost}=\frac{1}{2}(y_{1}+y_{2}). \eqno(14)
$$
We may write
$$
\tau=\frac{4p^{2}_{T}}{s} \cosh^2y^{*} \eqno(15)
$$
and
$$
x_{1}=\sqrt{\tau}e^{y_{boost}},~~~~
x_{2}=\sqrt{\tau}e^{-y_{boost}}.\eqno(16)
$$
\par
In our numerical calculation we adopt the MRS set G parton distribution
function $f_{g}(x_{i},Q^{2})$ \cite{ss1} and let the factorization scale
$Q^2=\hat{s}$. The numerical calculation is carried out around the
Tevatron and LHC energy ranges.

\vskip 5mm
{\Large{\bf III. Numerical Results and Discussions}}
\vskip 5mm
\par
In the numerical calculation we take the input parameters\cite{exp} as
$m_{b}=4.5 GeV$, $m_{c}=1.35 GeV$, $m_{t}=175 GeV$, $M_{W}=80.2226 GeV$,
$G_{F}=1.166392\times 10^{-5} (GeV)^{-2}$ and $\alpha=1/128$.
We adopt a simple one-loop formula for the running strong coupling
constant $\alpha_s$ as
$$
\alpha_s(\mu)=\frac{\alpha_{s}(m_Z)} {1+\frac{33-2 n_f} {6 \pi} \alpha_{s}
              (m_Z) \ln \left( \frac{\mu}{m_Z} \right) }. \eqno(14)
$$
where $\alpha_s(m_Z)=0.117$ and $n_f$ is the number of active flavors at
energy scale $\mu$.
\par
Figure 2 shows the cross sections for $g g \rightarrow t\bar{c} +
\bar{t}c$ as a function of the masses of the Higgs bosons $M_{s}$. The cross
sections are displayed with the three values of the gluon-gluon CMS energy
200 GeV, 400 GeV and 500 GeV respectively. Because there is no stringent bound
on the Higgs bosons masses, we choose $M_{s}$ in the range from 50 GeV to 800
GeV. The peak of each curve comes from s-channel resonance effects,
where $M_{s}=m_{h^{0}}=m_{A^{0}} \sim \sqrt{\hat{s}}$. From these curves we
find that the cross section can be obviously enhanced when $M_{s}$ gets
close to $\sqrt{\hat{s}}$.
\par
Figure 3 shows the cross sections of $g g \rightarrow t\bar{c}+
\bar{t}c$ as a function of $\sqrt{\hat{s}}$, and the three curves correspond
to the $M_{s}$ values 100 GeV, 250 GeV and 500 GeV, respectively.
For $M_{s}=100 GeV$, the effects of the widths of the Higgs bosons are not
obvious, and s-channel resonance effects are suppressed, since
$\sqrt{\hat{s}}$ is far beyond the Higgs boson masses $M_{s}$.
Therefore the curve of its cross section is relative flat with the increasing
of $\sqrt{\hat{s}}$. When $\sqrt{\hat{s}}$ approaches the value of $M_{s}$,
such as $M_{s}=250 GeV$ and $M_{s}=500 GeV$, the cross sections
will be enhanced by the s-channel resonance effects, and the width effects
become stronger, since the $h^0, A^0 \rightarrow t\bar{c}+\bar{t}c$
channels are opened. In Fig.3, we can see that the curve for $M_{s}=500 GeV$
shows a sharp peak around the position at $\sqrt{\hat{s}} \simeq 500 GeV$
due to the s-channel resonance effects and large width effects of $h^{0}$
and $A^{0}$.
\par
In figure 4 and figure 5 we show the transverse momentum spectrum of the
top quark at the Tevatron and LHC energies respectively, where we assume the
off-line analysis will require at least the event selection criterion of
involving one isolated high $p_{T}$ track with the cut of pseudorapidity
$\vert \eta\vert <2$. Again, the peaks on the curves
of $M_{s}=250$ and $M_{s}=500$ show the s-channel resonance effects,
where $M_{s}=m_{h^{0}}=m_{A^{0}} \sim \sqrt{\hat{s}}$.
\par
In figure 6 and 7 we show the cross section of $pp\rightarrow gg
\rightarrow t\bar{c}+\bar{t}c+X$ as a function of center-of-mass energy of
electron-positron system $\sqrt{s}$. The cross section
may reach 0.83 femtobarn when $M_{s}=100~GeV$ and $\sqrt{s}=2~TeV$ at the
Tevatron and 131 femtobarn when $M_{s}=100~GeV$ and $\sqrt{s}=14~TeV$ at the
LHC. For the Tevatron at 2 TeV we can expect about 4 raw events when
$M_{s}=100~GeV$ if we assume $5 fb^{-1}$ integrated luminosity, and for the LHC
at 14 TeV we can expect about $1.3\times 10^{4}$ raw events if we assume
$100 fb^{-1}$ integrated luminosity. Since the cross section of this
process is roughly scaled by $\lambda^{4}$, if we let $\lambda \simeq 1$,
the cross section will be 4 times larger. There are several potentially
severe backgrounds from the SM to the signal. A top quark with a mass
about $174GeV$, decays dominantly to $t\rightarrow W^{+}b$. In the $t\bar{c}
\rightarrow W^{+} b \bar{c}\rightarrow l^{+}\nu b \bar{c}$ detection mode,
the backgrounds are mainly from $t\bar{t} \rightarrow W+jets$ and
$t\bar{t} \rightarrow W W b\bar{b} \rightarrow l^{+} \nu q \bar{q}^{'} b
\bar{b}$ processes. The calculation shows that the cross section of top pair
production will reach about $5~pb$ at the Tevatron for $\sqrt{s}=1.8TeV$ and
about $10^{2}~pb$ at the LHC for $\sqrt{s}=14TeV$. If we use the $t\bar{c}+
\bar{t}c \rightarrow lepton+jets$ detection mode, the production rate of
$t\bar{c} \rightarrow W^{+} b \bar{c} \rightarrow l^{+}\nu b \bar{c}$ can
reach about $0.1~fb$ at the Tevatron and $14~fb$ at the LHC, while the possible
background from $t\bar{t} \rightarrow W W b\bar{b} \rightarrow l^{+} \nu
q \bar{q}^{'} b \bar{b}$ would be about $0.8~pb$ at the Tevatron and some dozens
picobarn at the LHC. The reduction of the
these backgrounds is possible through various kinematics cuts on the
transverse energy, on the rapidity of jets and leptons, or involving b-tagging.
Due to the very small production rate for the signal of $pp\rightarrow t\bar{c}
+\bar{t}c$, it is not so easy to suppress its backgrounds. Therefore the
further precise analyses are necessary to exclude these backgrounds.
\par
In summary, from our calculation, we can conclude that if we could efficiently
exclude the severe backgrounds of the $t\bar{c}(\bar{t}c)$ production signal,
it would be possible at the Tevatron and the LHC that the process
$p p \rightarrow g g \rightarrow t\bar{c}+\bar{t}c$ could be used to probe
the flavor changing interactions in the context of the THDM III.

\vskip 4mm
\noindent{\large\bf Acknowledgement:}
\par
This work was supported in part by the National Natural Science
Foundation of China(project numbers: 19675033, 19875049), the Youth Science
Foundation of the University of Science and Technology of China(USTC) and
a grant from the Research Fund for the Doctoral Program of Higher
Education(RFDP) of China.

\vskip 5mm
\noindent
{\Large{\bf Appendix}}
\vskip 5mm

We adopt the same definitions of one-loop A, B, C and D integral functions as
in Ref.\cite{abcd} and the references therein. The dimension $D=4- \epsilon$.
The integral functions are defined as
$$
A_{0}(m)=-\frac{(2\pi\mu)^{4-D}}{i\pi^{2}} \int d^{D}q \frac{1}{[q^2-m^2]} ,
$$
$$
\{B_{1};B_{\mu};B_{\mu\nu}\}(p,m_1,m_2) =
\frac{(2\pi\mu)^{4-D}}{i\pi^{2}} \int d^{D}q
\frac{\{1;q_{\mu};q_{\mu\nu}\}}{[q^2-m_{1}^{2}][(q+p)^2-m_{2}^{2}]} ,
$$
$$
\{C_{0};C_{\mu};C_{\mu\nu};C_{\mu\nu\rho}\}(p_1,p_2,m_1,m_2,m_3) =
-\frac{(2\pi\mu)^{4-D}}{i\pi^{2}}
$$
$$
\times \int d^{D}q
\frac{\{1;q_{\mu};q_{\mu\nu};q_{\mu\nu\rho}\}}
{[q^2-m_{1}^{2}][(q+p_1)^2-m_{2}^{2}][(q+p_1+p_2)^2-m_{3}^{2}]} ,
$$
$$
\{D_{0};D_{\mu};D_{\mu\nu};D_{\mu\nu\rho};D_{\mu\nu\rho\alpha}\}
(p_1,p_2,p_3,m_1,m_2,m_3,m_4) =
\frac{(2\pi\mu)^{4-D}}{i\pi^{2}}
$$
$$
\times \int d^{D}q \{1;q_{\mu};q_{\mu\nu};q_{\mu\nu\rho};
q_{\mu\nu\rho\alpha}\}
$$
$$
\times \{[q^2-m_{1}^{2}][(q+p_1)^2-m_{2}^{2}][(q+p_1+p_2)^2-m_{3}^{2}]
[(q+p_1+p_2+p_3)^2-m_{4}^{2}]\}^{-1}.
$$

In our calculation we take the strange quark mass $m_{s}=0$.
The $M^{\hat{s}_{1}}$ and $M^{\hat{s}_{2}}$ in equation (9) can
be written as

$$
\begin{array}{lll}
M^{\hat{s}_{1}}&=&
\frac{i\alpha^{2}_{s} g^{2}}{16 \pi^{2} M_{W}^{2}} m_{t}\sqrt{m_{t}m_{c}}
\epsilon_{\mu}(p_{3})\epsilon_{\nu}(p_{4})
\bar{u}(p_{1})\\
&&\cdot\{
2 a_{h^{0}} m_{t}^2 (C_{0} + 4 C_{22} - 4 C_{23})[p_{3}, -p_{1}
 - p_{2}, m_{t}, m_{t}, m_{t}](p_{1}^{\mu}p_{1}^{\nu}+p_{1}^{\mu}p_{2}^{\nu}+
 p_{1}^{\nu}p_{2}^{\mu}+p_{2}^{\mu}p_{2}^{\nu})\\
&& + 2a_{h^{0}}m_{t}^2 g^{\mu\nu} (B_{0}[-p_{1} - p_{2}, m_{t}, m_{t}] -
       ((p_{1}+p_{2}) \cdot p_{3} C_{0}
      +4 C_{24} )[p_{3}, -p_{1} - p_{2}, m_{t}, m_{t}, m_{t}]\\
 &&      + 2 i a_{A^{0}} m_{t}^2 C_{0}[p_{3}, -p_{1} - p_{2}, m_{t}, m_{t}, m_{t}]
      \epsilon^{\mu\nu\alpha\beta}\gamma_{5}
        (p_{1}^{\alpha}p_{3}^{\beta}+p_{2}^{\alpha}p_{3}^{\beta})
        \}v(p_{2}),
\end{array}
\eqno{(A.1)}
  $$
where
$$a_{h^{0}}=\frac{1}{\hat{s}-m_{h^{0}}^2+i\Gamma_{h^{0}}m_{h^{0}}},~~~~~
  a_{A^{0}}=\frac{1}{\hat{s}-m_{A^{0}}^2+i\Gamma_{A^{0}}m_{A^{0}}}.$$

$$
\begin{array}{lll}
M^{\hat{s}_{2}}&=&
\frac{i\alpha^{2}_{s} g^{2}}{128 \pi^{2} M_{W}^{2}\hat{s}} m_{t}\sqrt{m_{t}m_{c}}
\epsilon_{\mu}(p_{3})\epsilon_{\nu}(p_{4})
\bar{u}(p_{1})\\
&&\cdot \{ - 16 m_{t} (C_{11} -C_{12} + C_{21}- C_{23})
[-p_{1}, p_{1} + p_{2}, M_{s}, m_{t}, m_{t}]
(p_{1}^{\mu}p_{2}^{\nu}-p_{1}^{\nu}p_{2}^{\mu}) \\
&& + (8 ( 2 C_{24}  + m_{t}^2 ( C_{0} - C_{11} + C_{12} - C_{21} - C_{22} + 2 C_{23})
  - 2  (p_{1}\cdot p_{2}) (C_{22}+C_{23}))\\
&&  [-p_{1}, p_{1} + p_{2}, M_{s}, m_{t}, m_{t}]+
4 m_{t}^{2} B_{1}[-p_{1},m_{t}, M_{s}]-\\
&&4 m_{t} m_{c} B_{1}[p_{2}, m_{t}, M_{s}]
+4 m_{t}^{2} B_{1}[-p_{1}, m_{b}, M_{s}]+\\
&&4 m_{t} m_{c} B_{1}[p_{2}, m_{b}, M_{s}]+ m_{t}^{2} B_{1}[-p_{1},m_{t}, M_{s}]\\
&&- m_{t} m_{c} B_{1}[p_{2}, m_{t}, M_{s}]- m_{t}^{2} B_{1}[-p_{1}, m_{b}, M_{s}]\\
&&+ m_{c} m_{t} B_{1}[p_{2}, m_{b}, M_{s}] )
  (\gamma^{\nu}p_{1}^{\mu}-\gamma^{\mu}p_{1}^{\nu}+\gamma^{\nu}p_{2}^{\mu}
  -\gamma^{\mu}p_{2}^{\nu} +g^{\mu\nu}\rlap/p_{3})\\
&&+(-4 ( 2 m_{t} C_{24} - 2 m_{t}^3 C_{0}
 + m_{t} (m_{t}^2+2 (p_{1}\cdot p_{2})-4 (p_{1}\cdot p_{3}))
    (C_{11}-C_{12}+C_{21}+C_{22}\\
 &&   - 2 C_{23}) + 4 (p_{2}\cdot p_{3}) (m_{c} C_{12} -m_{t} C_{22} +m_{t} C_{23}))
 [-p_{1}, p_{1} + p_{2}, M_{s}, m_{t}, m_{t}] \\
&& - (m_{c} m_{t}^2 C_{0}
 + 2 m_{t} C_{24} + m_{t}^3 (C_{11}  - C_{12}  + C_{21}  + C_{22}  - 2 C_{23})
  + 2 m_{t} ((p_{1}\cdot p_{2})-\\
&&  2 (p_{1}\cdot p_{2})) (C_{11}
  - C_{12} + C_{21} + C_{22} - 2 C_{23})
  + 4 (p_{2}\cdot p_{3}) (m_{c} C_{12} \\
&&  + m_{t} (C_{23} -C_{22}))
  [-p_{1}, p_{1} + p_{2}, M_{s}, m_{b}, m_{b}]
 -m_{t}^{3}  B_{1}[-p_{1},m_{t}, M_{s}] +\\
&& m_{t}^{2} m_{c} B_{1}[p_{2}, m_{t}, M_{s}]
 - m_{t}^{3} B_{1}[-p_{1}, m_{b}, M_{s}] +(m_{c} m_{t}^{2}+\\
&& 2 m_{c} p_{1}\cdot p_{2} - 4 m_{c} p_{1}\cdot p_{3}) B_{1}[p_{2}, m_{b}, M_{s}])
 g^{\mu\nu} -4 m_{t} (C_{11}\\
 && -C_{12}+C_{21}-C_{23})
[-p_{1}, p_{1} + p_{2}, M_{s}, m_{b}, m_{b}]
(p_{1}^{\mu}p_{2}^{\nu}-p_{1}^{\nu}p_{2}^{\mu}
+\gamma_{5}p_{1}^{\nu}p_{2}^{\mu}- \gamma_{5}p_{1}^{\mu}p_{2}^{\nu}) \\
&&+( 2 (2 C_{24} +  m_{c} m_{t} C_{0} -  m_{t}^2 (C_{11}-C_{12}+ C_{21} + C_{22} -2 C_{23})
  - 2(p_{1}\cdot p_{2}) (C_{22}-C_{23}))\\
 && [-p_{1}, p_{1} + p_{2}, M_{s}, m_{b}, m_{b}]+
 4 m_{t}^{2} B_{1}[-p_{1}, m_{b}, M_{s}]-\\
&& 4 m_{c} m_{t} B_{1}[p_{2}, m_{b}, M_{s}])
  (\gamma^{\nu}p_{1}^{\mu}-\gamma^{\mu}p_{1}^{\nu}+\gamma^{\nu}p_{2}^{\mu}
  -\gamma^{\mu}p_{2}^{\nu} +g^{\mu\nu}\rlap/p_{3})
- (m_{c} m_{t}^2 C_{0} - 2 m_{t} C_{24}\\
&&  - m_{t}^3 (C_{11}  - C_{12}  + C_{21}  + C_{22}  - 2 C_{23})
 - 2 m_{t} ((p_{1}\cdot p_{2})-2 (p_{1}\cdot p_{2})) (C_{11}-C_{12}+C_{21} \\
&&  + C_{22} - 2 C_{23})
  + 4 (p_{2}\cdot p_{3}) (m_{c} C_{12} - m_{t} (C_{23} -C_{22}))
  [-p_{1}, p_{1} + p_{2}, M_{s}, m_{b}, m_{b}]\gamma_{5} g^{\mu\nu}\\
&&+(( 4 C_{24} - 2 m_{c} m_{t} C_{0}
  - 2 m_{t}^2 (C_{11} - C_{12} + C_{21}  +C_{22} - 2 C_{23})
  + 4 (p_{1}\cdot p_{2}) (C_{23}\\
&&  -C_{22})) [-p_{1}, p_{1} + p_{2}, M_{s}, m_{b}, m_{b}])+
2 m_{t}^{2} B_{1}[-p_{1}, m_{b}, M_{s}]- \\
&& 2 m_{c} m_{t} B_{1}[p_{2}, m_{b}, M_{s}])
\gamma_{5}( \gamma^{\nu}p_{1}^{\mu}-\gamma^{\mu}p_{1}^{\nu}
+\gamma^{\nu}p_{2}^{\mu}-\gamma^{\mu}p_{2}^{\nu}-g^{\mu\nu}\rlap/p_{3})
 \} v(p_{2}),
\end{array}
\eqno{(A.2)}
  $$
The amplitude of $M^{\hat{t}}$ can be written as
$$
\begin{array}{lll}
M^{\hat{t}}&=&
\frac{i\alpha_{s}^{2} g^{2}}{64 \pi^{2} M_{W}^{2}} m_{t}\sqrt{m_{t}m_{c}}
\epsilon_{\mu}(p_{3})\epsilon_{\nu}(p_{4}) \bar{u}(p_{1})
 (f_{1}p_{1}^{\mu}p_{1}^{\nu}+f_{2}p_{1}^{\mu}p_{2}^{\nu}+
f_{3}p_{1}^{\nu}p_{2}^{\mu}
+f_{4}p_{2}^{\mu}p_{2}^{\nu}\\
&&+ f_{5}\gamma^{\nu}p_{1}^{\mu}+f_{6}\gamma^{\mu}p_{1}^{\nu}
+ f_{7}\gamma^{\nu}p_{2}^{\mu}
+f_{8}\gamma^{\mu}p_{2}^{\nu}
+ f_{9}\gamma^{\mu}\gamma^{\nu}+f_{10}\gamma^{\nu}\gamma^{\mu}
+ f_{11}\rlap/p_{3}p_{1}^{\mu}p_{1}^{\nu}\\
&&+f_{12}\rlap/p_{3}p_{1}^{\mu}p_{2}^{\nu}
+ f_{13}\rlap/p_{3}p_{1}^{\nu}p_{2}^{\mu}
+f_{14}\rlap/p_{3}p_{2}^{\mu}p_{2}^{\nu}
+ f_{15}\rlap/p_{3}\gamma^{\nu}p_{1}^{\mu}
+f_{16}\rlap/p_{3}\gamma^{\mu}p_{1}^{\nu}
+ f_{17}\rlap/p_{3}\gamma^{\nu}p_{2}^{\mu}\\
&&+f_{18}\rlap/p_{3}\gamma^{\mu}p_{2}^{\nu}
+ f_{19}\rlap/p_{3}\gamma^{\mu}\gamma^{\nu}
+f_{20}\rlap/p_{3}\gamma^{\nu}\gamma^{\mu}
+ f'_{1}\gamma_{5} p_{1}^{\mu}p_{1}^{\nu}
+f'_{2}\gamma_{5} p_{1}^{\mu}p_{2}^{\nu}+
f'_{3}\gamma_{5} p_{1}^{\nu}p_{2}^{\mu}\\
&&+f'_{4}\gamma_{5} p_{2}^{\mu}p_{2}^{\nu}+
f'_{5}\gamma_{5} \gamma^{\nu}p_{1}^{\mu}
+f'_{6}\gamma_{5} \gamma^{\mu}p_{1}^{\nu}
+ f'_{7}\gamma_{5} \gamma^{\nu}p_{2}^{\mu}
+f'_{8}\gamma_{5} \gamma^{\mu}p_{2}^{\nu}
+ f'_{9}\gamma_{5} \gamma^{\mu}\gamma^{\nu}\\
&&+f'_{10}\gamma_{5} \gamma^{\nu}\gamma^{\mu}
+ f'_{11}\gamma_{5} \rlap/p_{3}p_{1}^{\mu}p_{1}^{\nu}
+f'_{12}\gamma_{5} \rlap/p_{3}p_{1}^{\mu}p_{2}^{\nu}
+ f'_{13}\gamma_{5} \rlap/p_{3}p_{1}^{\nu}p_{2}^{\mu}
+f'_{14}\gamma_{5} \rlap/p_{3}p_{2}^{\mu}p_{2}^{\nu}\\
&&+ f'_{15}\gamma_{5} \rlap/p_{3}\gamma^{\nu}p_{1}^{\mu}
+f'_{16}\gamma_{5} \rlap/p_{3}\gamma^{\mu}p_{1}^{\nu}
+ f'_{17}\gamma_{5} \rlap/p_{3}\gamma^{\nu}p_{2}^{\mu}
+f'_{18}\gamma_{5} \rlap/p_{3}\gamma^{\mu}p_{2}^{\nu}\\
&&+ f'_{19}\gamma_{5} \rlap/p_{3}\gamma^{\mu}\gamma^{\nu}
+f'_{20}\gamma_{5} \rlap/p_{3}\gamma^{\nu}\gamma^{\mu} )v(p_{2}),
\end{array}
\eqno{(A.3)}
$$

where the $f_{i}s$ and $f'_{i}s$ are expressed explicitly as,

$$
\begin{array}{lll}
f'_{1}& =&
 4(m_{c} (D_{38}-D_{310}) + m_{t} (D_{32} - D_{36}))
     [p_{3}, -p_{1}, -p_{2}, m_{b}, m_{b}, M_{s}, m_{b}]
\end{array}
\eqno{(A.4)}
$$

$$
\begin{array}{lll}
f_{1}& =&
f'_{1}(m_{t}\rightarrow -m_{t})+
  8 m_{t} (D_{11} -D_{12} + D_{21} - D_{24} - D_{25} + D_{26})
      [p_{1}, -p_{3}, -p_{4}, M_{s}, m_{t}, m_{t}, m_{t}]
\end{array}
\eqno{(A.5)}
$$

$$
\begin{array}{lll}
f'_{2}&=& 4 a_{1} (m_{c} C_{22} - m_{t} C_{12} - m_{t} C_{23})
     [-p_{4}, p_{2}, m_{b}, m_{b}, M_{s}]\\
&&     + 4(m_{c} (D_{39}-D_{310}) - m_{t} (D_{22} - D_{26} + D_{36} - D_{38}))
     [p_{3}, -p_{1}, -p_{2}, m_{b}, m_{b}, M_{s}, m_{b}]
\end{array}
\eqno{(A.6)}
$$

$$
\begin{array}{lll}
f_{2}& =&
f'_{2}(m_{t}\rightarrow -m_{t})+
 8 a_{1} m_{t} C_{11}[-p_{2}, p_{4}, M_{s}, m_{t}, m_{t}]\\
&&     -8 m_{t} (D_{12}+D_{24}- D_{26})
      [p_{1}, -p_{3}, -p_{4}, M_{s}, m_{t}, m_{t}, m_{t}]
\end{array}
\eqno{(A.7)}
$$

$$
\begin{array}{lll}
f'_{3}&=& 4 (m_{c} (D_{39}-D_{37}) - m_{t} (D_{310}- D_{38}))
     [p_{3}, -p_{1}, -p_{2}, m_{b}, m_{b}, M_{s}, m_{b}]
\end{array}
\eqno{(A.8)}
$$

$$
f_{3} =
f'_{3}(m_{t}\rightarrow -m_{t}) -8 m_{t} (D_{25}-D_{26})
    [p_{1}, -p_{3}, -p_{4}, M_{s}, m_{t}, m_{t}, m_{t}]
\eqno{(A.9)}
$$

$$
f'_{4} = 4 (m_{c} (D_{33} - D_{37}) + m_{t} (D_{23} - D_{26} - D_{310} \\
+ D_{39})) [p_{3}, -p_{1}, -p_{2}, m_{b}, m_{b}, M_{s}, m_{b}]
\eqno{(A.10)}
$$

$$
f_{4} =
f'_{4}(m_{t}\rightarrow -m_{t})+
 8 m_{t} D_{26}[p_{1}, -p_{3}, -p_{4}, M_{s}, m_{t}, m_{t}, m_{t}]
\eqno{(A.11)}
$$

$$
\begin{array}{lll}
f'_{5}&=&- 2 a_{2} a_{3} m_{t}^{2} B_{1}[-p_{1}, m_{b}, M_{s}]
     - 2 a_{1} a_{2} (m_{t}^{2} - 2 (p_{1}\cdot p_{3})) B_{1}
     [-p_{1} + p_{3}, m_{b}, M_{s}]\\
&&     - 2 a_{1} a_{3} m_{c} m_{t} B_{1}[p_{2}, m_{b}, M_{s}]
     + 2 a_{2} ( m_{c} m_{t} (C_{0} + 2 C_{11} + C_{21})- 2 C_{24} )
     [p_{1}, -p_{3}, M_{s}, m_{t}, m_{t}]\\
&&   + 2 a_{1} (-2 C_{24} - m_{b}^{2} C_{0} +m_{c}^{2} ( C_{22} - 2 C_{23}) -
    m_{c} m_{t} C_{12} + m_{t}^{2} (C_{11} + C_{21})\\
&&    + 2 (p_{1}\cdot p_{2}-p_{2}\cdot p_{3}) (C_{11}
  - C_{12}+ C_{21} - C_{23}) \\
&&  - 2 (p_{1}\cdot p_{3}) (C_{11} + C_{21}))
  [-p_{4}, p_{2}, m_{b}, m_{b}, M_{s}]\\
&&  + 2 (2 D_{27} + 2 D_{312} + m_{b}^{2} D_{0} -m_{c}^{2} D_{23}
  - m_{c} m_{t} (D_{22}- D_{26})\\
&&      + 2 (p_{2}\cdot p_{3}) (D_{25}-D_{26}))
      [p_{3}, -p_{1}, -p_{2}, m_{b}, m_{b}, M_{s}, m_{b}]
\end{array}
\eqno{(A.12)}
$$

$$
\begin{array}{lll}
f_{5}& =&
f'_{5}(m_{t}\rightarrow -m_{t})+
     4 a_{2} a_{3} m_{t}^{2} B_{0}[-p_{1}, m_{t}, M_{s}]\\
&&      +4 a_{1} a_{2} m_{t}^{2} B_{0} [-p_{1} + p_{3}, m_{t}, M_{s}]
     - 4 a_{1} a_{3} m_{t}^{2} B_{0} [p_{2}, m_{t}, M_{s}]\\
&&     - 4 a_{2} m_{t}^{2} (C_{0}+C_{11})[p_{1}, -p_{3}, M_{s}, m_{t}, m_{t}]
     - 4 a_{1} m_{t}^{2} [-p_{2}, p_{4}, M_{s}, m_{t}, m_{t}]\\
&&    +4 m_{t}^{2} (D_{0} +D_{11} - D_{13})
    [p_{1}, -p_{3}, -p_{4}, M_{s}, m_{t}, m_{t}, m_{t}]
\end{array}
\eqno{(A.13)}
$$

$$
\begin{array}{lll}
f'_{6}&=& 2 (4 (D_{311} - D_{312}) + m_{b}^{2} (D_{11} - D_{12})
   - m_{c}^{2} (D_{37} - D_{39}) \\
&&   + m_{t}^{2} (D_{32} - D_{36})
   + 2 (p_{1}\cdot p_{2}) (D_{38}- D_{310})
  - 2 (p_{1}\cdot p_{3}) (D_{22}-D_{24}-D_{34}+D_{36})\\
&&  + 2 (p_{1}\cdot p_{2}) (D_{35}- D_{310}))
  [p_{3}, -p_{1}, -p_{2}, m_{b}, m_{b}, M_{s}, m_{b}]
\end{array}
\eqno{(A.14)}
$$

$$
f_{6}=f'_{6}
\eqno{(A.15)}
$$

$$
\begin{array}{lll}
f'_{7}&=&
2(-4 D_{313} - m_{b}^{2} D_{13} + m_{c}^{2} D_{33} + m_{c} m_{t} D_{23} +
  m_{t}^{2} (D_{26} + D_{38}) + 2 (p_{1}\cdot p_{2}) (D_{23} +D_{39})\\
&&  - 2 (p_{1}\cdot p_{3}) (D_{25} +D_{310}+D_{23}+ D_{37}))
  [p_{3}, -p_{1}, -p_{2}, m_{b}, m_{b}, M_{s}, m_{b}]
\end{array}
\eqno{(A.16)}
$$

$$
f_{7} =
f'_{7}(m_{t}\rightarrow -m_{t})
 -4 m_{t}^{2} D_{13}[p_{1}, -p_{3}, -p_{4}, M_{s}, m_{t}, m_{t}, m_{t}]
\eqno{(A.17)}
$$

$$
\begin{array}{lll}
f'_{8}&=& -4 a_{1} (p_{1}\cdot p_{3}) (C_{12} + C_{23})
     [-p_{4}, p_{2}, m_{b}, m_{b}, M_{s}]\\
&&     + 2 (2 (D_{27} + 2 D_{311} + D_{313}) + m_{b}^{2} (D_{0}
    + D_{11}) - m_{c}^{2} (D_{23}+ D_{37})\\
&&    - m_{t}^{2} (D_{22}+ D_{36})
    [p_{3}, -p_{1}, -p_{2}, m_{b}, m_{b}, M_{s}, m_{b}]\\
&&   -  (p_{1}\cdot p_{2}) (D_{26} +D_{310})
  + 2 (p_{1}\cdot p_{3}) (D_{12} - D_{13}+ 4 D_{24}- D_{26} + D_{34})\\
&&  + 2 (p_{2}\cdot p_{3}) (D_{25} +D_{35}))
  [p_{3}, -p_{1}, -p_{2}, m_{b}, m_{b}, M_{s}, m_{b}]
\end{array}
\eqno{(A.18)}
$$

$$
f_{8}= f'_{8}
\eqno{(A.19)}
$$

$$
\begin{array}{lll}
f'_{9}&=&-  a_{2} a_{3} m_{t}^{3} B_{1}[-p_{1}, m_{b}, M_{s}]\\
&&    +2 a_{1} a_{2} m_{c} (p_{1}\cdot p_{3}) B_{1}
       [-p_{1} + p_{3}, m_{b}, M_{s}]
     - a_{2} m_{t} ( 2 C_{24} \\
&&     - m_{t}^{2} (C_{0} + 2 C_{11} + C_{21})+
       2 (p_{1}\cdot p_{3}) (C_{0}+C_{11}+C_{12}+C_{23}))
       [p_{1}, -p_{3}, M_{s}, m_{t}, m_{t}]\\
&&     -2 a_{1} m_{c} (p_{1}\cdot p_{3}) C_{12}
     [-p_{4}, p_{2}, m_{b}, m_{b}, M_{s}]
    +  (4 m_{c} (D_{313} \\
&&    + m_{b}^{2} D_{13}  - m_{c}^{2} D_{33}) +
     m_{t} (2 D_{27} + 4 D_{312} + m_{b}^{2} (D_{0} + D_{12})
     - m_{c}^{2} (D_{23}+ D_{39})\\
&&    - m_{t} (m_{c} D_{38} - m_{t} D_{22}
    - m_{t} D_{32})) - 2 (p_{1}\cdot p_{2}) (m_{c} D_{39} \\
&&    + m_{t} D_{26}+ m_{t} D_{38})
    + 2 (p_{1}\cdot p_{3}) (m_{c} D_{12} - m_{c} D_{13}
    + m_{c} D_{310}+m_{t} D_{24} + m_{t} D_{36})\\
&&  +2 (p_{2}\cdot p_{3}) (m_{c} D_{37} + m_{t} D_{25}
  + m_{t} D_{310}))[p_{3}, -p_{1}, -p_{2}, m_{b}, m_{b}, M_{s}, m_{b}]
\end{array}
\eqno{(A.20)}
$$

$$
\begin{array}{lll}
f_{9}& =&
f'_{9}(m_{t}\rightarrow -m_{t})
 - 2 a_{2} a_{3} m_{t}^{3} B_{0}[-p_{1}, m_{t}, M_{s}]\\
&&   - 4 a_{1} a_{2} m_{t} (p_{1}\cdot p_{3}) B_{0}[-p_{1} + p_{3}, m_{t}, M_{s}]\\
&&   +4 a_{2} m_{t} (p_{1}\cdot p_{3}) C_{0}[p_{1}, -p_{3}, M_{s}, m_{t}, m_{t}]\\
&&   +4 a_{1} m_{t} (p_{1}\cdot p_{3}) C_{0}[-p_{2}, p_{4}, M_{s}, m_{t}, m_{t}]\\
&&   +4 m_{t} (D_{27}+ D_{311} - D_{313})
     [p_{1}, -p_{3}, -p_{4}, M_{s}, m_{t}, m_{t}, m_{t}]
\end{array}
\eqno{(A.21)}
$$

$$
f'_{10}= -2 (m_{c} D_{313}+ m_{t} D_{312})[p_{3}, -p_{1}, -p_{2}, m_{b}, m_{b}, M_{s}, m_{b}]
\eqno{(A.22)}
$$

$$
f_{10} =
f'_{10}(m_{t}\rightarrow -m_{t})
 -4 m_{t} D_{27}[p_{1}, -p_{3}, -p_{4}, M_{s}, m_{t}, m_{t}, m_{t}]
\eqno{(A.23)}
$$

$$
f'_{11}= 4 (D_{22} - D_{24} - D_{34} + D_{36})[p_{3}, -p_{1}, -p_{2}, m_{b}, m_{b}, M_{s}, m_{b}]
\eqno{(A.24)}
$$

$$
f_{11}=f'_{11}
\eqno{(A.25)}
$$

$$
f'_{12}= 4 (D_{13}-D_{12}  - 2 D_{24} + 2 D_{26} + D_{310} - D_{34})[p_{3}, -p_{1}, -p_{2}, m_{b}, m_{b}, M_{s}, m_{b}]
\eqno{(A.26)}
$$

$$
f_{12}= f'_{12}
\eqno{(A.27)}
$$

$$
f'_{13}
\eqno{(A.28)}
$$

$$
f_{13}= f'_{13}
\eqno{(A.29)}
$$

$$
f'_{14}= 4 (D_{23} - D_{25} - D_{35} + D_{37})[p_{3}, -p_{1}, -p_{2}, m_{b}, m_{b}, M_{s}, m_{b}]
\eqno{(A.30)}
$$

$$
f_{14}=f'_{14}
\eqno{(A.31)}
$$

$$
\begin{array}{lll}
f'_{15}&=& - 2 a_{2} m_{t} (C_{11} - C_{12} + C_{21} - C_{23})
      [p_{1}, -p_{3}, M_{s}, m_{t}, m_{t}]\\
&&      +2(m_{c} (D_{13}-D_{12}+ D_{26}) + m_{t} (D_{22} - D_{24}))
      [p_{3}, -p_{1}, -p_{2}, m_{b}, m_{b}, M_{s}, m_{b}]
\end{array}
\eqno{(A.32)}
$$

$$
\begin{array}{lll}
f_{15} &=&
f'_{15}(m_{t}\rightarrow -m_{t})+
4 a_{2} m_{t} C_{11}[p_{1}, -p_{3}, M_{s}, m_{t}, m_{t}]\\
&&-4 m_{t} (D_{11}- D_{13}) [p_{1}, -p_{3}, -p_{4}, M_{s}, m_{t}, m_{t}, m_{t}]
\end{array}
\eqno{(A.33)}
$$

$$
f'_{16}= 2 (m_{c} (D_{25} - D_{26}) - m_{t} (D_{22} - D_{24}))
     [p_{3}, -p_{1}, -p_{2}, m_{b}, m_{b}, M_{s}, m_{b}]
\eqno{(A.34)}
$$

$$
f_{16} =
f'_{16}(m_{t}\rightarrow -m_{t})+
 4 m_{t} (D_{11} - D_{12})[p_{1}, -p_{3}, -p_{4}, M_{s}, m_{t}, m_{t}, m_{t}]
\eqno{(A.35)}
$$

$$
f'_{17}= 2 (m_{c} D_{23} - m_{t} (D_{25} - D_{26}))
       [p_{3}, -p_{1}, -p_{2}, m_{b}, m_{b}, M_{s}, m_{b}]
\eqno{(A.36)}
$$

$$
f_{17} =
f'_{17}(m_{t}\rightarrow -m_{t})+
 4 m_{t} D_{13}[p_{1}, -p_{3}, -p_{4}, M_{s}, m_{t}, m_{t}, m_{t}]
\eqno{(A.37)}
$$

$$
\begin{array}{lll}
f'_{18}&=& 2 a_{1} (m_{c} C_{22} - m_{t} C_{12} - m_{t} C_{23})
      [-p_{4}, p_{2}, m_{b}, m_{b}, M_{s}]\\
&&      +2 (m_{c} (D_{25}-D_{23}) + m_{t} (D_{12} - D_{13} + D_{24} - D_{26}))
      [p_{3}, -p_{1}, -p_{2}, m_{b}, m_{b}, M_{s}, m_{b}]
\end{array}
\eqno{(A.38)}
$$

$$
\begin{array}{lll}
f_{18} &=&
f'_{18}(m_{t}\rightarrow -m_{t})+
 4 a_{1} m_{t} C_{11}[-p_{2}, p_{4}, M_{s}, m_{t}, m_{t}]\\
&&      -4 m_{t} D_{12} [p_{1}, -p_{3}, -p_{4}, M_{s}, m_{t}, m_{t}, m_{t}]
\end{array}
\eqno{(A.39)}
$$

$$
\begin{array}{lll}
f'_{19}&=&-  a_{2} a_{3} m_{t}^{2} B_{1}[-p_{1}, m_{b}, M_{s}]\\
&&     -  a_{1} a_{2} (m_{t}^{2} - 2 (p_{1}\cdot p_{3})) B_{1}
     [-p_{1} + p_{3}, m_{b}, M_{s}]\\
&&    -  a_{1} a_{3} m_{c} m_{t} B_{1}[p_{2}, m_{b}, M_{s}]
     - a_{2} (2 C_{24} - m_{t} (m_{c} C_{0} + m_{t} C_{11}\\
&&     +m_{t} C_{21}) + 2 (C_{12}+C_{23}) (p_{1}\cdot p_{3}))
    [p_{1}, -p_{3}, M_{s}, m_{t}, m_{t}]\\
&&    +  a_{1} (-2 C_{24} - m_{b}^{2} C_{0} +m_{c}^{2} ( C_{22} - 2 C_{23}) -
     m_{c} m_{t} C_{12} \\
&&     + m_{t}^{2} (C_{11} + C_{21})
     + 2 ((p_{1}\cdot p_{2})-(p_{2}\cdot p_{3})) (C_{11} - C_{12}
    + C_{21}- C_{23})\\
&&    - 2 (p_{1}\cdot p_{3}) (C_{11} + C_{21}))
    [-p_{4}, p_{2}, m_{b}, m_{b}, M_{s}]\\
&&  + (4 (D_{27} + D_{311}) + 2 m_{b}^{2} (D_{0} + D_{11})
  - 2 m_{c}^{2} (D_{23} + D_{37})
 - D_{13} m_{c} m_{t} \\
&& - m_{t}^{2} (D_{12} + D_{22}+ D_{36})
 - 2 (p_{1}\cdot p_{2}) (D_{12} + 2 D_{26}+ D_{310})
  + 2 (p_{1}\cdot p_{3}) (D_{12} \\
&&  + 2 D_{24}+ D_{34}) + 2 (p_{2}\cdot p_{3}) (D_{13} + 2 D_{25}+ D_{35}))
 [p_{3}, -p_{1}, -p_{2}, m_{b}, m_{b}, M_{s}, m_{b}]
\end{array}
\eqno{(A.40)}
$$

$$
\begin{array}{lll}
f_{19} &=&
f'_{19}(m_{t}\rightarrow -m_{t})+
 2 a_{2} a_{3} m_{t}^{2} B_{0}[-p_{1}, m_{t}, M_{s}]\\
&&      +2 a_{1} a_{2} m_{t}^{2} B_{0} [-p_{1} + p_{3}, m_{t}, M_{s}]
     - 2 a_{1} a_{2} m_{t}^{2} B_{0} [p_{2}, m_{t}, M_{s}]\\
&&     - 2 a_{2} m_{t}^{2} C_{0}[p_{1}, -p_{3}, M_{s}, m_{t}, m_{t}]
     - 2 a_{1} m_{t}^{2} C_{0}[-p_{2}, p_{4}, M_{s}, m_{t}, m_{t}]\\
&&     + 2 m_{t}^{2} D_{0} [p_{1}, -p_{3}, -p_{4}, M_{s}, m_{t}, m_{t}, m_{t}]
\end{array}
\eqno{(A.41)}
$$

$$
f'_{20}= -2 D_{311} [p_{3}, -p_{1}, -p_{2}, m_{b}, m_{b}, M_{s}, m_{b}]
\eqno{(A.42)}
$$

$$
f_{20}= f'_{20}
\eqno{(A.43)}
$$

where
$$a_{1}=\frac{1}{\hat{t}-m_{t}^{2}},~~
a_{2}=\frac{1}{\hat{t}-m_{c}^{2}}
~~and ~~a_{3}=\frac{1}{m_{t}^{2}-m_{c}^{2}}.$$

$$ M^{\hat{u}}=M^{\hat{t}} ~~~(p_{3}\leftrightarrow p_{4}, ~~\mu \leftrightarrow
\nu, \hat{t}\leftrightarrow \hat{u})~~~~~~~(A.44)$$

\vskip 10mm
\begin{flushleft} {\bf Figure Captions} \end{flushleft}

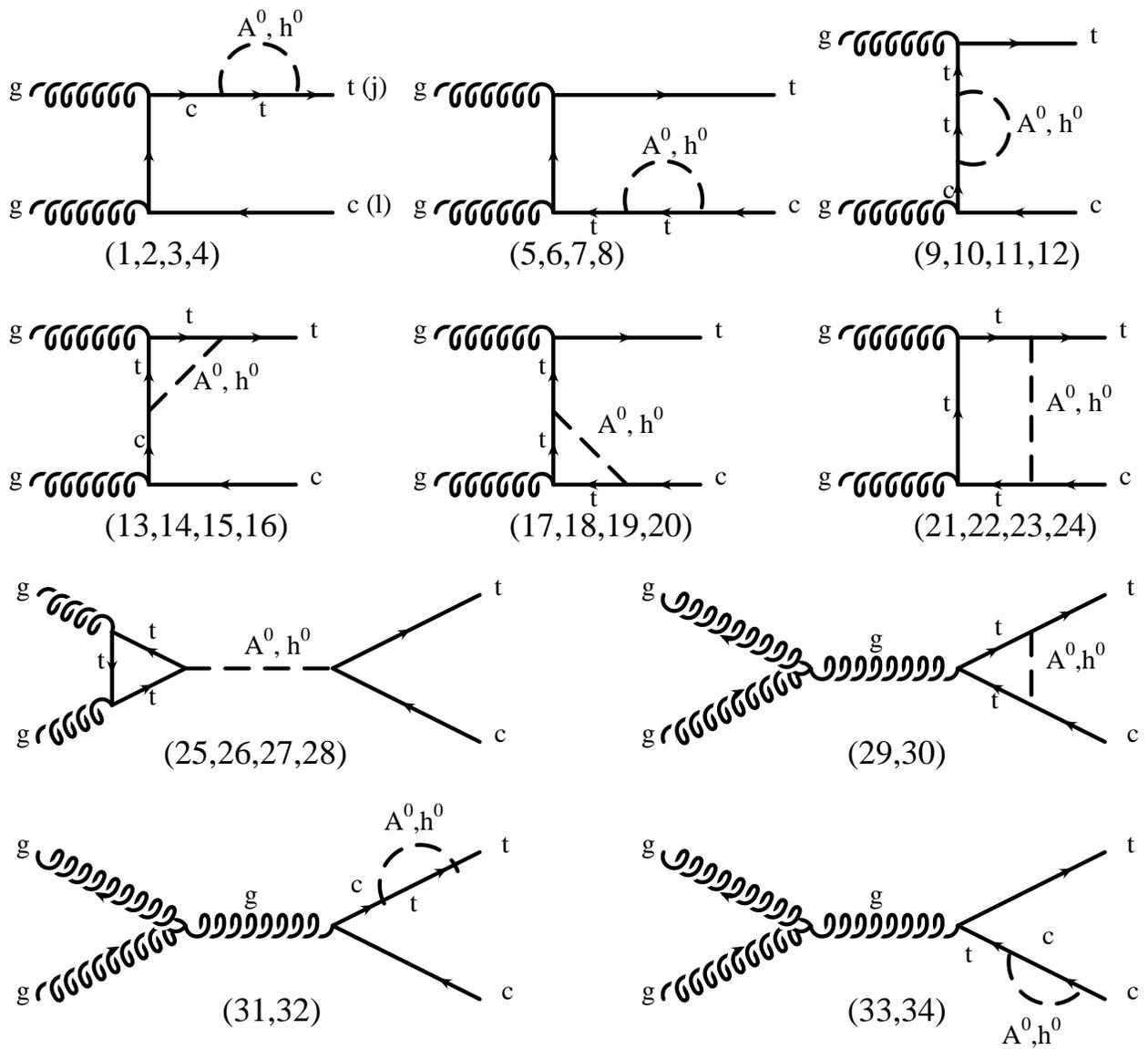
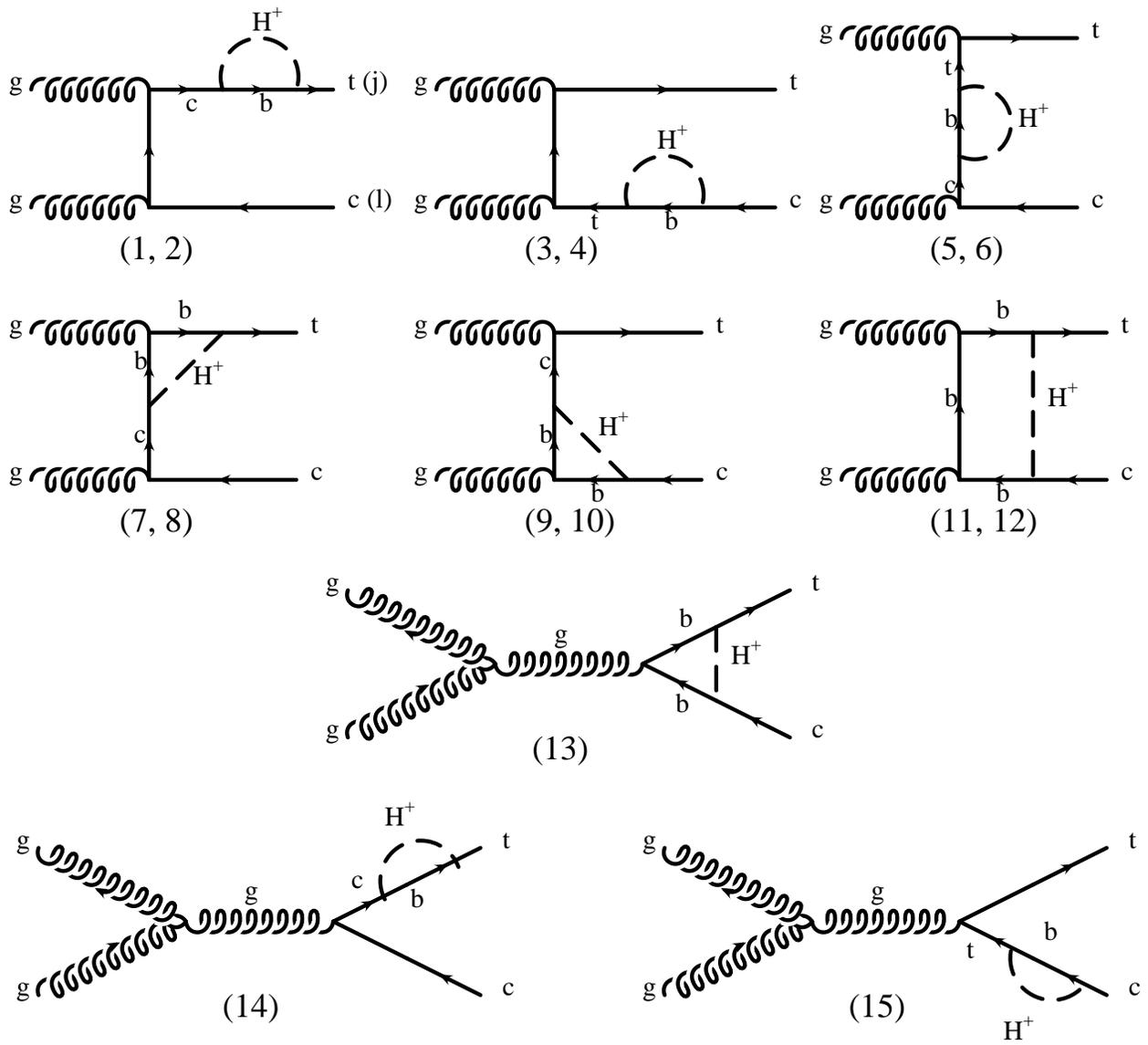
{\bf Fig.1}(a)(b) The Feynman diagrams of the subprocess $g g \rightarrow
t\bar{c}$.

{\bf Fig.2} Total cross sections of the subprocess $g g \rightarrow
t\bar{c} + \bar{t}c$ as function of $M_{s}$. The solid curve is for
$\sqrt{\hat{s}}=200 GeV$, the dashed curve is for
$\sqrt{\hat{s}}=400 GeV$ and the dotted curve is for
$\sqrt{\hat{s}}=500 GeV$.

{\bf Fig.3} Total cross sections of the subprocess $g g \rightarrow
t\bar{c} + \bar{t}c$ as function of $\sqrt{\hat{s}}$. The solid curve is for
$M_{s}=100~GeV$, the dashed curve is for $M_{s}=250~GeV$ and the
dotted curve is for $M_{s}=500~GeV$.

{\bf Fig.4} The transverse momentum spectrum of top quark for the Tevatron
at 2 TeV. The solid curve is for $M_{s}=100~GeV$, the dashed curve is for
$M_{s}=250~GeV$ and the dotted curve is for $M_{s}=500~GeV$.

{\bf Fig.5} The transverse momentum spectrum of top quark for the LHC
at 14 TeV. The solid curve is for $M_{s}=100~GeV$, the dashed curve is for
$M_{s}=250~GeV$ and the dotted curve is for $M_{s}=500~GeV$.

{\bf Fig.6} Total cross sections of the process $p p\rightarrow
g g \rightarrow t\bar{c} + \bar{t}c$ as function of $\sqrt{s}$ at the Tevatron.
The solid curve is for $M_{s}=100~GeV$, the dashed curve is for
$M_{s}=250~GeV$ and the dotted curve is for $M_{s}=500~GeV$.

{\bf Fig.7} Total cross sections of the process $p p\rightarrow
g g \rightarrow t\bar{c} + \bar{t}c$ as function of $\sqrt{s}$ at the LHC.
The solid curve is for $M_{s}=100~GeV$, the dashed curve is for
$M_{s}=250~GeV$ and the dotted curve is for $M_{s}=500~GeV$.

\end{large}
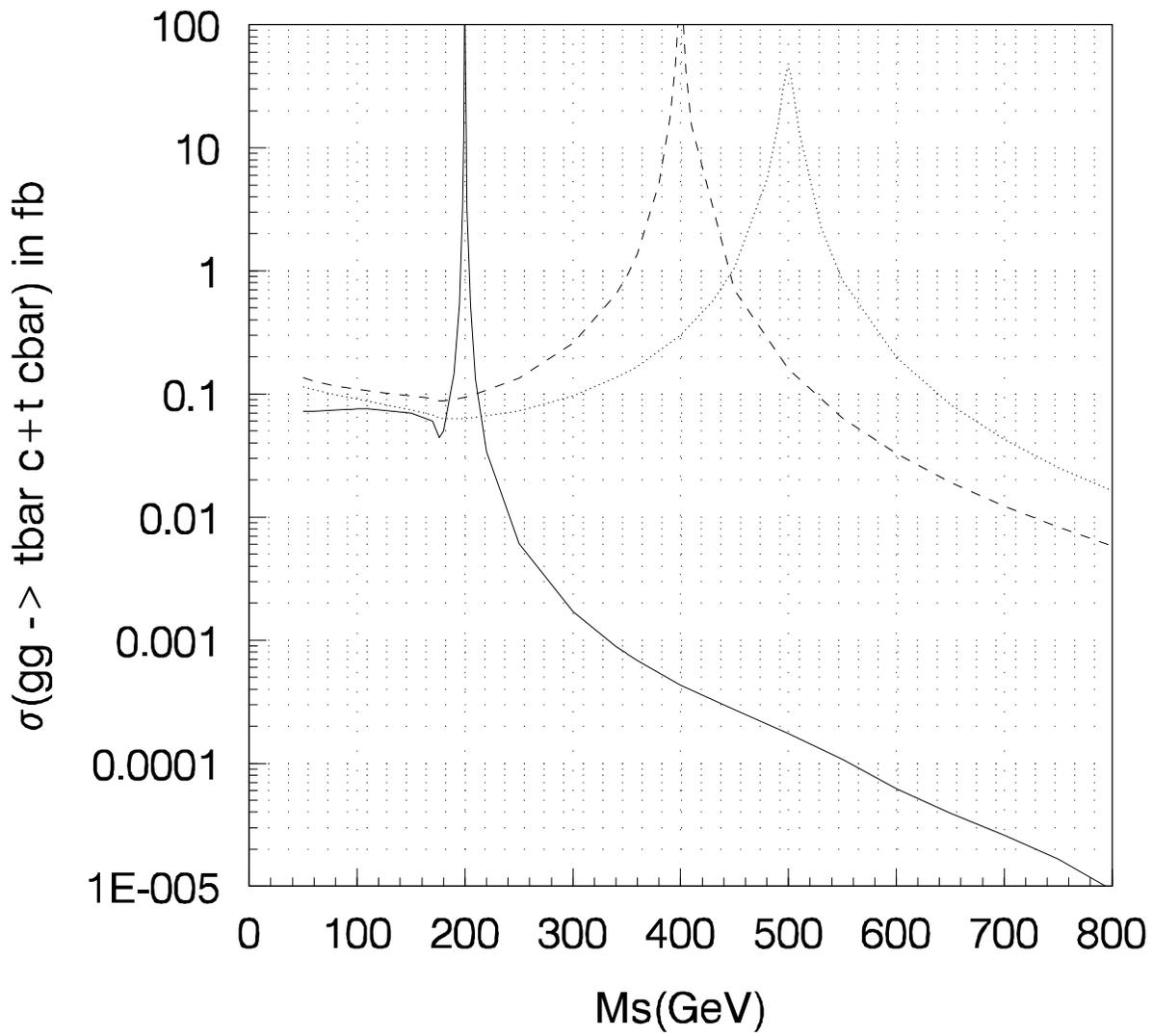
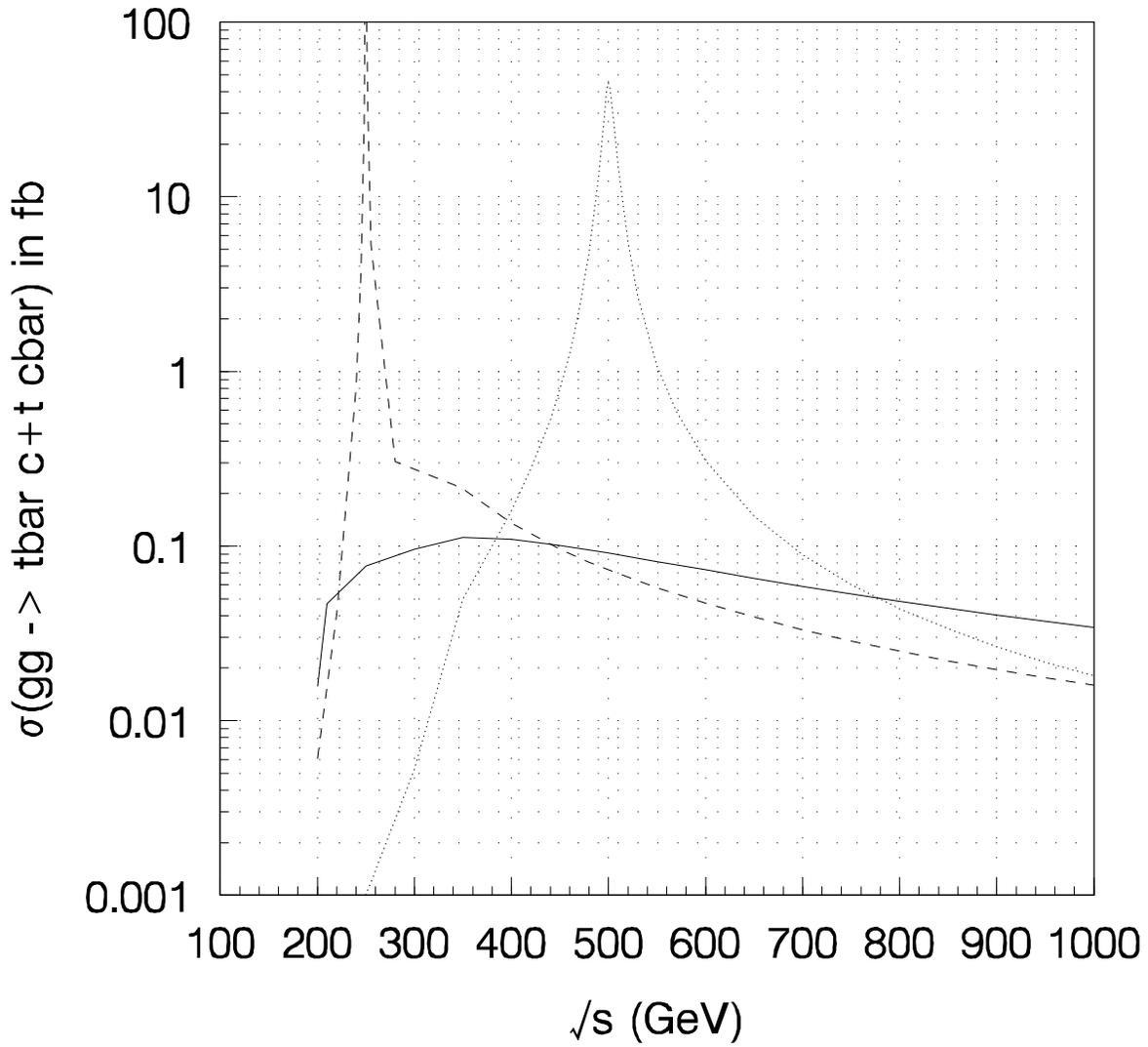
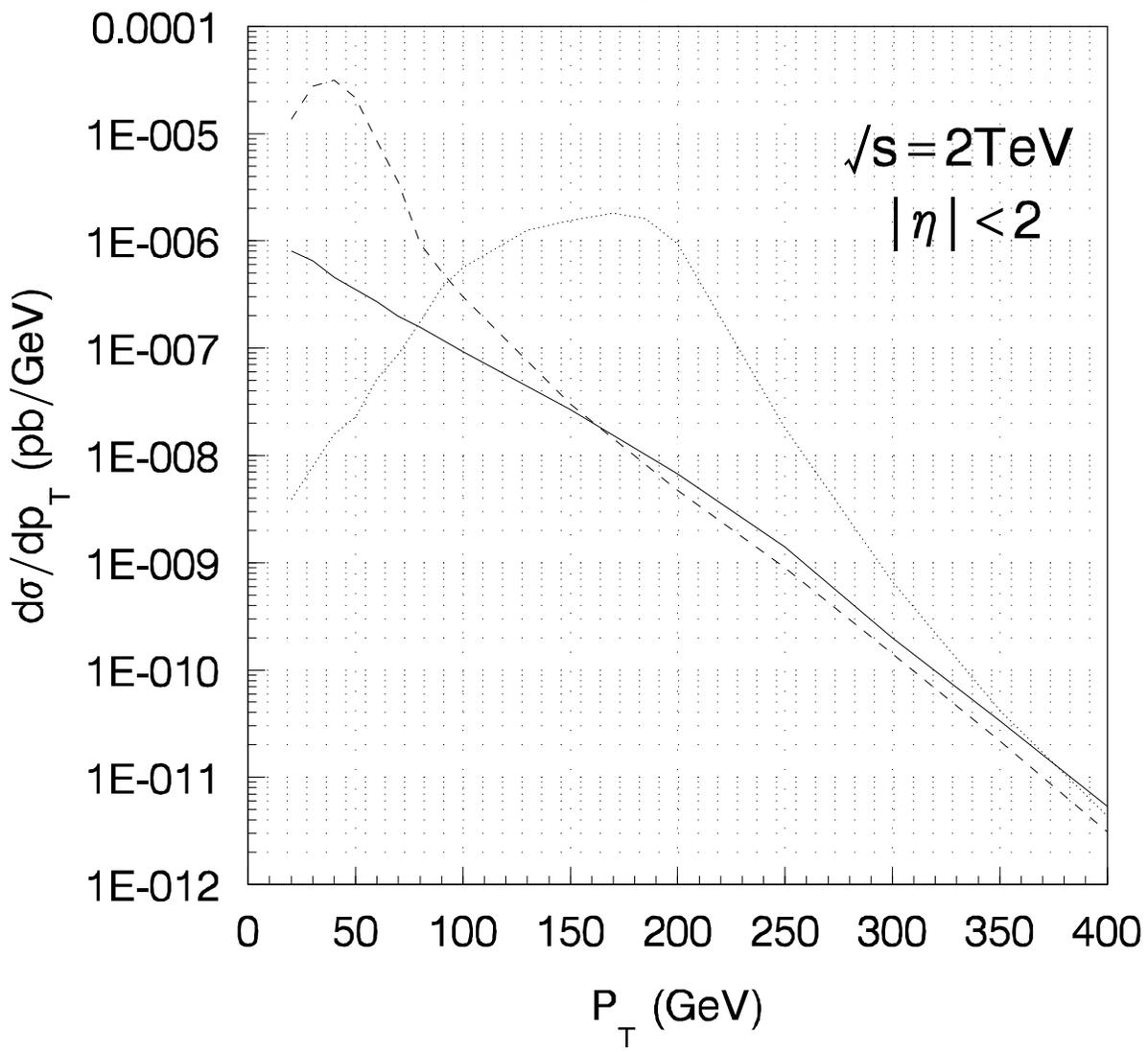
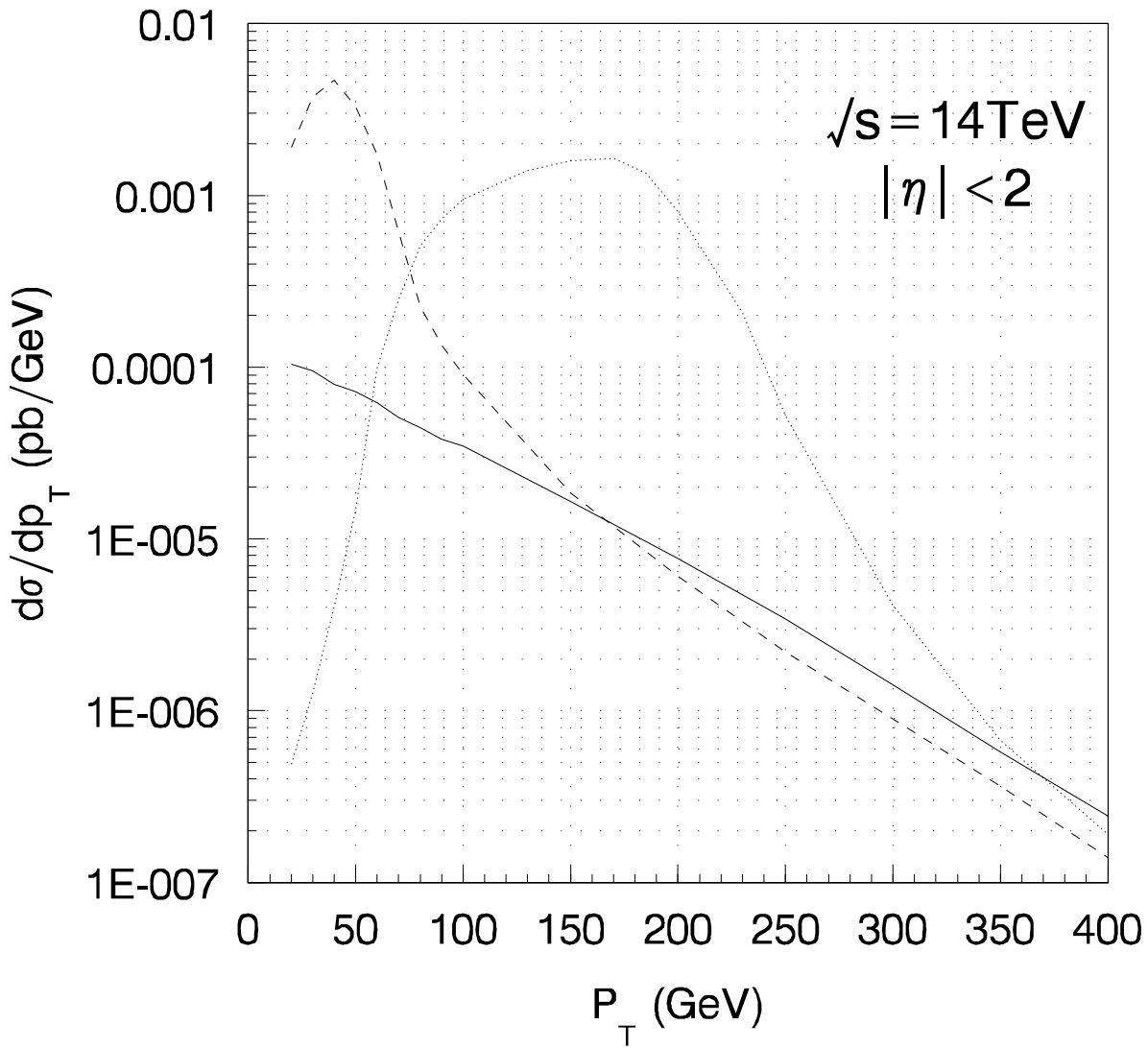
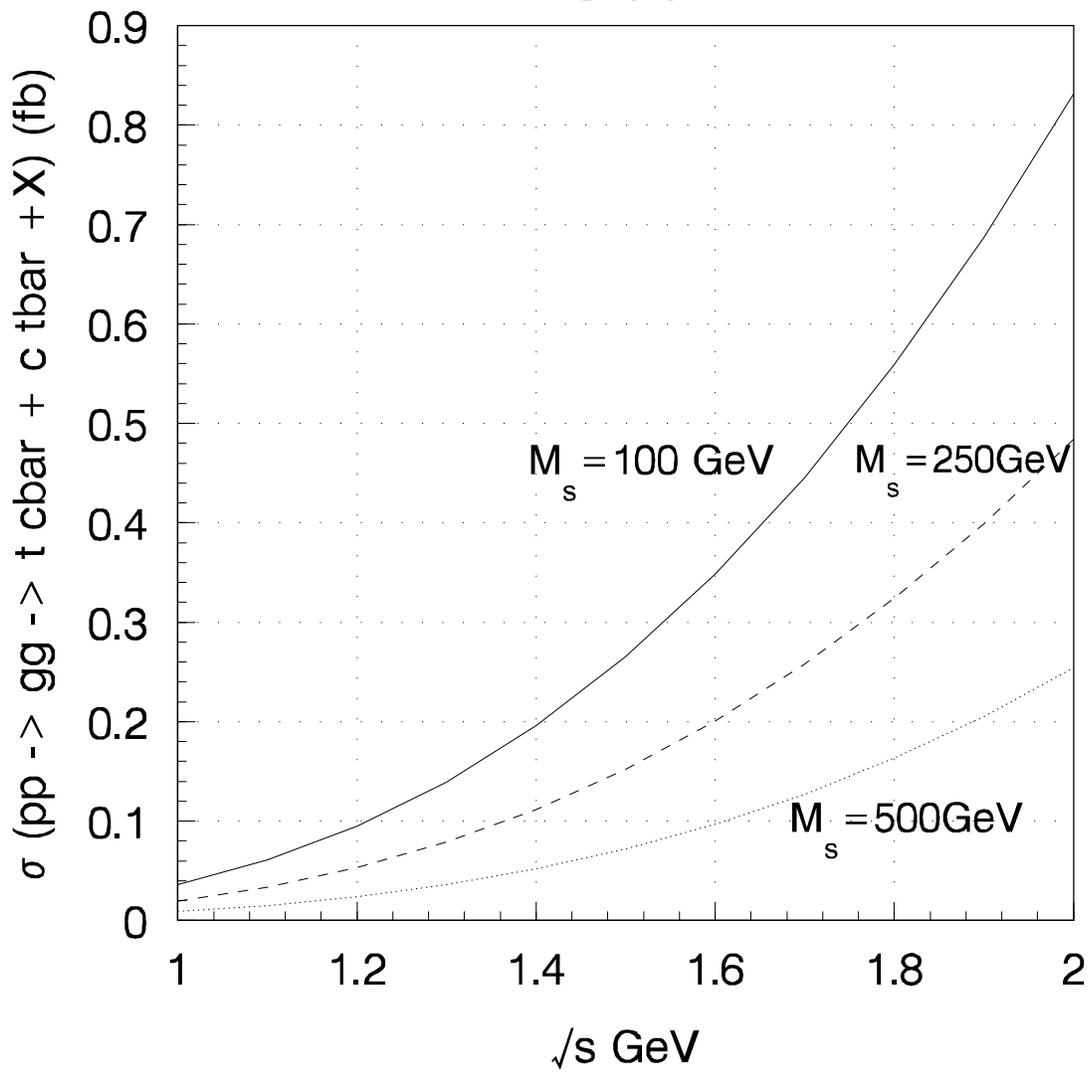
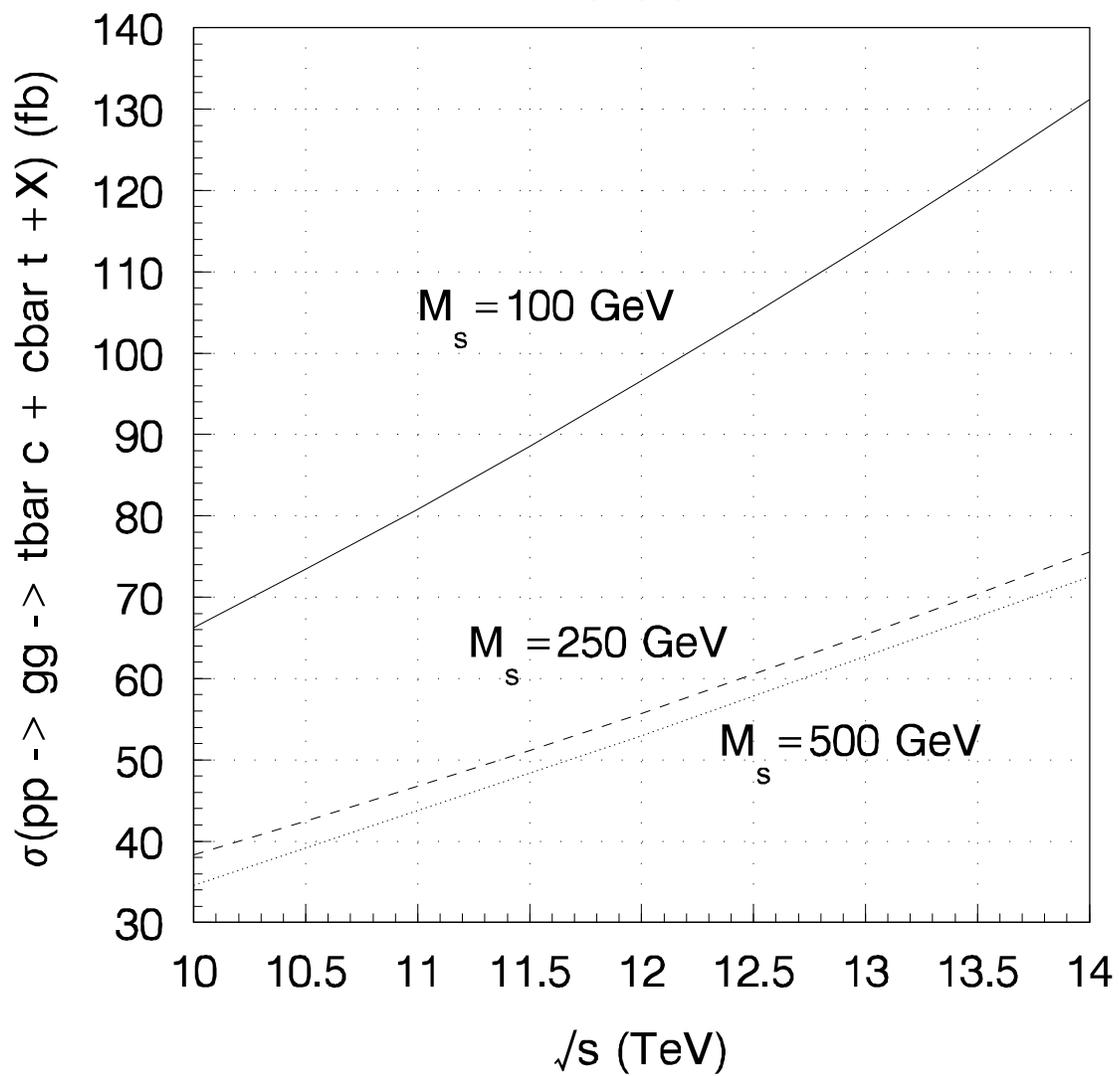
\end{document}